# In Quest for proper Mediums for Technology Transfer in Software Engineering

Florian Grigoleit, Antonio Vetrò, Daniel Méndez Fernández, Wolfgang Böhm
Technische Universität München
Garching bei München, Germany
{grigolei,vetro, mendezfe, boehmw }@in.tum.de

Philipp Diebold
Fraunhofer IESE
Fraunhofer-Platz 1, Kaiserslautern, Germany
Philipp.diebold@iese.fraunhofer.de

*Abstract—Background*: Successful transfer of the results of research projects into practice is of great interest to all project participants. It can be assumed that different transfer mediums fulfill technology transfer (TT) with different levels of success and that they are impaired by different kinds of barriers. *Objective*: The goal of this study is to gain a better understanding about the different mediums used for TT in software engineering, and to identify barriers weakening the success of the application of such mediums. *Method*: We conducted an exploratory study implemented by a survey in the context of a German research project with a broad range of used mediums. *Results*: The main reported barriers were low expectations of usefulness, no awareness of existence, lack of resources, or inadequateness in terms of outdated material or being in an immature state. *Conclusions:* We interpreted our results as symptoms of a lack of a dissemination plan in the project. Further work will be needed to explore the implications for the transfer of research results (knowledge and techniques) to practice.

I. INTRODUCTION

Publicly funded research projects are instances of broader intervention programs whose expected benefits respond to the needs of particular regions or fields. Funding agencies evaluate the results and impacts of the funded projects to determine whether they reflect the initial objectives and have addressed the initial needs. In most cases, it is out of the scope of single projects to evaluate the long-term effects of the developed innovations on broad aspects corresponding to the addressed goals, e.g. the economic growth of a particular region or industrial field. However, the achievement of such goals depends on the success of the individual research projects. Thus, it is important to understand when a research project can be considered successful. In computer science, particularly in industry-related disciplines like Software Engineering (SE), the success of a project can be twofold:

(1) Producing technically sound solutions addressing the original objectives

(2) Transferring the results to the academic community, to project partners, as well as to external stakeholders to foster innovation.

This paper focuses on the second part. More specifically, we want to get a first understanding about what typical barriers to the use of specific mediums for the dissemination of results are. This allows, in the long run, to further explore means to support the effective dissemination of project results. The underlying assumption is that a successful dissemination of results is very dependent on, beyond other things, the choice of proper mediums and that we can increase the success of the Technology Transfer (TT) by the choice of such mediums.

A vital part of TT is therefore the used set of transfer mediums. A transfer medium can be any kind of communication medium used for transferring information from the transferor to the transferee. In our previous work [2] we provided a classification schema [3] that allocates a set of mediums to different abstraction levels of a taxonomy of models for TT in Software Engineering [1]. Examples of mediums are wikis (which belongs to the Communication Model), personnel exchanges (People-mover Model), consultancy programs (Vendor Model). Considering the variety of mediums in aspects of *abstraction*, *format*, *contents*, and *typical use*, it is intuitive to assume that different mediums are used with different purposes and might achieve the success and effectiveness differently. In [2], we reported that commonly used mediums are human-intensive, i.e. meetings, workshops, and similar. Considering this and the common reliance on artifact-intensive mediums as well as the role of dissemination as one of the keys for successful transfer of research results, it is of great importance to identify the barriers to the successful use of transfer mediums.

We conduct our exploration in the ambit of a large German research project, SPES-XT[1], with a consortium of more than 20 partners from academia and industry (representing corporations key for the German automotive, automation, and avionic sector) developing modeling techniques for embedded systems [6]. Their project results were disseminated by a large variety of mediums, ranging from workshops over guidelines to a project wiki. In this paper, we extend our previous work by assessing the barriers impairing the use of mediums and the reaching of their purposes. To this end, we conducted a study on the effectiveness of project dissemination and dissemination mediums.

The reminder of the paper is organized as follows: Section II covers the design of the study with a particular focus on the instrument used. Section III explains the data analysis procedures and a classification for transfer mediums. In Sect. IV we present the analysis methodology, followed by the results of the study (Sec. V) and their discussion (Sec. VI). Finally we conclude our paper in Sect. VII.

---

[1] Please refer to our website for information on the project and the partners: http://spes2020.informatik.tu-muenchen.de/spes_xt-home.html

## II. STUDY DESIGN

### A. Goals and research questions

The goal of our study is to elicit the barriers impairing the use of transfer mediums as well as their success.

To reach this goal, we define the following research questions (RQs).

- *RQ1*: What were the barriers for accessing and using mediums?
- *RQ2*: What were the barriers impairing the successful achievement of the purpose for using the mediums?

We decided to investigate our research questions via survey research [8]. In the following, we introduce the instrument used while details on the overall survey can be taken from our previously published material [2], [3].

### B. Instrument Design

As mediums, we generally distinguish between artifacts, like publications, and events, e.g. workshops, due to our observations in [1] concerning the preferred usage of human intensive mediums. Exemplary mediums are website, wiki, mailing list, guidelines [9], workshops, or summer schools. Events like workshops were broken down to their specific instances (e.g. Envision 2013 and 2014) to avoid losing valuable information (two different workshops may get different answers). To answer our research questions, we formulate subsequent questions in the instrument:

- *Med01 - Which artifact / events did you use / attend?* (Nominal, Multiple options from the list of mediums used in SPES-XT)
- *Med02 / 03 - What was the purpose of using / attending the artifact / event?* (Open text)
- *Med04 - Did the artifact / event reach its purpose?* (Nominal: Yes/Partially/No)
- *Med05 - Why did the artifact / event not or only partially reach its purpose?* (Open text)
- *Med06 - For which reason didn't you use/attend these artifacts / events?* (Nominal, Multiple options: Too much effort required/ Not useful enough/ Lack of competencies/ Refusal from management/Others)
- *Med07 - For which other reason didn't you use those artifacts / events?* (Open text)

The participant pre-selected a subset of used mediums in question *Med01* and then received the according questions *Med02-05,* covering only the selected mediums. The subset of mediums presented in *Med05* includes only artifacts/events with answers "Partially" or "No" in *Med04*. Question *Med06* is answered on each unused medium (from question *Med01*), and question *Med07* only from items answered as "Others" in *Med06*. Considering the results from [2], we expect that certain, less frequently used mediums create barriers making them less attractive for potential users. To identify such barriers and further obstacles, *RQ2* is refined into the questions asked in *Med06 and Med07*.

In brief, the answers to questions *Med01-04* build the set of mediums that are evaluated in questions *Med05-07*, which are directly connected to the research questions RQ1 (*Med06 / Med07*) and RQ2 (*Med05*).

## III. DATA ANALYSIS METHODOLOGY

We categorize both artifacts and events in three types, described in Table I and Table II. The answers of *Med01* are not taken into account in the data analysis, because they only served as a filter to the following questions (see Sect. II). Regarding the other questions, we apply Grounded Theory for all open questions and free text options, and code the answers following the procedure described in [7].

In *Med02-03*, participants were asked about the purpose of using the selected mediums: the codes for the different purposes are listed and explained in Table III.

TABLE I: ARTIFACT CATEGORIES

| Category | Description | Instances |
|---|---|---|
| Technical Artifacts | Artifacts suitable or even designed as basis or to support development activities | Building Blocks, Scenarios, Guidelines |
| Web-based Artifacts | Web-based Artifacts containing on various aspects on the project and the results | Blog on TT, SPES Website, SPES Wiki |
| Academic / Teaching Artifacts | Artifacts generally intended for academic of teaching purposes | Courses and Projects, Lecture Material, Surveys, Publications |

TABLE II: EVENT CATEGORIES

| Category | Description | Instances |
|---|---|---|
| Internal Dissemination Activities | Dissemination Activities only intended for inner-organizational purposes | IDAs Industry, IDAs Academic Partners |
| Project Events | SPES project events | SPES Summer School, SPES 202 Conclusion Event |
| Conferences | Conferences and Workshops for SPES-related topics | EITEC'14, ENVISION'13, ENVISION'14, SWORDS'14 |

TABLE III: CODING FOR MEDIUMS PURPOSE (MED02-03)

| Code | Description | Exemplary statement |
|---|---|---|
| Medium used as **information** source | Medium is used to obtain unspecified information | "Information about SPES methods", "find out what is going on" |
| Medium used to support **development** | Medium is used as a basis for development or testing | "Structuring the development process", "application on methods and techniques" |
| Medium used for **internal dissemination** | Medium is used to disseminate artifacts within ones organization | "Internal presentation", "needed for transfer in the praxis" |
| Medium used for **external dissemination** | Medium is used to disseminate results external of SPES | "Publication", "presentation of DSE techniques and methods" |
| Medium used for **communication** | Medium is explicitly used for communication | "Scientific exchange", "all day communication" |

## IV. RESULTS

### A. Descriptive data

The questionnaire was online for five weeks in March and April 2015; we obtained 28 completed questionnaires, four working in academic research, 16 in industry. Eight participants did not state their affiliation. We cannot provide the respondents rate, because the survey invitations were spread via mailing lists.

Based on the answers to *Med02*, we observe that technical artifacts are mostly used to support developments activities. Only guidelines are mostly used for internal dissemination. Web-based artifacts are, in general, used for information retrieval. Artifacts with an academic purpose, such as publications, are used either for internal dissemination or for information retrieval. Conferences are mostly used for external dissemination, the internal dissemination activities (IDAs) are used for the dissemination among the project partners (in SPES-XT), and the project events are finally used for general information retrieval. As the data doesn't indicate any other purpose or target group, we assume that the participants themselves mainly make use of the mediums listed. Further, we observe that IDAs are used not only for internal dissemination, but also for practical development tasks. The workshops are used also for communication and not only for external dissemination.

### B. Answer to research questions

In the following, we summarize our results structured according to our research questions

We report the results in the following tables:

- Table IV shows the barriers to usage (RQ1), coded from *Med 07*.
- Table V shows the barriers for success (achievement of the purpose, RQ2)

In addition, we segment results by mediums types and purpose, in the following tables:

- Table VI reports most cited barriers by medium types
- Table VII reports most cited barriers by medium purpose

TABLE IV: BARRIERS TO USAGE – RQ1

| Code | Description | Exemplary statement | Frequency |
|---|---|---|---|
| Medium **not required** | Participant does not require the present medium | "already sufficiently knowledgeable" | 33 |
| Medium **immature** | Medium does not offer the required quality or maturity | "Weren't mature", "information not ready" | 4 |
| **No resources** for use of Medium | Organization doesn't offer resources for using the medium | "Lack of manpower" | 24 |
| Medium **not available** | No access to medium | "Was not invited" | 8 |
| Medium **not suitable** for participant | Participation / use not possible for participant due to affiliation | "Not in academia", "not suitable for Bosch" | 16 |
| Medium **unknown** to participant | Medium unknown to the user | "didn't know that this existed" | 13 |

TABLE V: BARRIERS TO ACHIEVEMENT OF PURPOSE – RQ2

| Code | Description | Exemplary statement | Frequency |
|---|---|---|---|
| Medium **not relevant** | Medium not relevant, due to its content or the role of the participant | "No relevant information available", "information too specific" | 7 |
| Medium **insufficient** | Medium does not offer the standard or quality required | "Need to be refined", "quality of deliverables sometimes not sufficient" | 7 |
| Medium **not applicable** | Use of medium for the intended purpose not possible or limited | "Not always applicable", "integration of new methods into product dev. process is sometimes difficult / expensive" | 3 |
| Medium **outdated** | Medium not always up to date | "Update frequency too low", "not up to date" | 3 |
| Medium **incomplete** | Medium is not ready for regular use because of missing information | "Not yet available", "missing information" | 3 |

TABLE VI: BARRIERS CLASSIFIED BY MEDIUM TYPES

| Medium Type | Most frequently stated barrier for using Medium (RQ1) | Most frequently stated barrier for achievement of purpose (RQ2) | Frequency RQ1/RQ2 |
|---|---|---|---|
| Technical Artifacts | Immature | Incomplete | 4/3 |
| Web-based Artifacts | Not required | Outdated | 7/2 |
| Academic Artifacts | Not required | Not relevant | 6/1 |
| Internal Dissemination Activities | No resource | Insufficient | 4/1 |
| Project Events | Not suitable | - | 9/- |
| Conferences | No resource | Not relevant | 16/3 |

TABLE VII: BARRIERS CLASSIFIED BY MEDIUM PURPOSE

| Medium Purpose | Most frequently stated barrier for using Medium (RQ1) | Most frequently stated barrier for achievement of purpose (RQ2) |
|---|---|---|
| Development | Immature / not required | Insufficient / incomplete |
| Information | Not required / unknown | Outdated / not relevant |
| Communication | Not required / no resources | Not relevant |
| Internal Dissemination | No resources / not required | Insufficient / not Applicable |
| External Dissemination | Not required / no resources | Not relevant |

## V. DISCUSSION

Analyzing the answer to the barriers to usage (RQ1), we observe that technical and web-based artifacts, thus completely artifact-intensive mediums [2], are often unused because of their perceived insufficiency or lack of usefulness. This implies either that the expectations on artifacts are often too high, or that produced artifacts were of insufficient quality. For web-based artifacts, the barrier "unknown" was stated recurrently: for instance, for the SPES_XT-TT blog, almost all participants stated that they had no prior knowledge about it. This indicates that the general communication of existing mediums is also an important barrier.

We observe the general tendency that most of the artifact-based mediums were defined as either incomplete or insufficient. More specifically, for the technical artifacts, the answers indicate that some of the artifacts did not meet the expected quality standards. The same applies to the web-based artifacts, where mostly relevance and update frequency were criticized. For conferences, several participants stated that the information presented was not relevant for their purposes. The answers to the artifacts with a more academic purpose and the internal dissemination activities do not indicate any major issues, but their success was not fully achieved (see RQ2).

Finally, the events were deemed as "not required" and the lack of attendance was justified with insufficient resources. Since our previous investigation [2] indicates a general preference on human-intensive mediums over artifact-intensive mediums, we conclude that either the focus of the events was the actual barrier or the way the events were announced and executed.

Regarding the barriers to the successful achievement of mediums' purpose (RQ2), results indicate that the most frequently mentioned barriers, among the ones we proposed, were "too much effort required" and "not useful enough". However, many respondents selected the option "others", providing their own explanations. In general, we observe that the technical artifacts did not achieve their purpose either because they were "not relevant" to the participants or their "content was not required", or because they were "not considered mature enough". Also, a few participants stated that they had "no prior knowledge of the artifacts". The web-based artifacts were mainly not used because they were "not relevant to the participants" (an exception here is the blog on technology transfer, which was unknown to several participants, as found in RQ1 too). The academic purpose artifacts were mostly not used because they were "not required" or "not considered suitable for the intended purposes". A number of participants from industry also stated that the conferences were not suitable for them, because they were not in academia. Finally, the internal dissemination activities were usually not attended, because they were "limited to specific organizations" and other participants could not participate or did not have information about the event.

To summarize our discussion, the first lesson learned from this analysis is that participants in our research projects did not access or use a medium mainly because of low perception of usefulness, no awareness of existence and lack of resources (for events). The second lesson learned was that medium's usage did not imply achievement of its purpose, mainly for perceived irrelevance, or inadequateness in terms of outdated material or being in an immature state.

In general, we observed that dissemination activities have been not reserved the proper attention and resources, and this could endanger the successful transfer of knowledge and techniques. This translates into a pragmatic take away for our next research projects, in which a dissemination plan will be made since the beginning of the project and usage and appropriateness of the mediums will be continuously evaluated.

## VI.  CONLCUSION AND FUTURE WORK

In this short paper, we explored on the barriers of technology transfer mediums in a large German research project involving both academic and industry partners. Our experience revealed that the participants to the research project under study made partially use of the mediums for the transfer of research results (in terms of knowledge or techniques), with low achievement of the purpose of using them. The main reported barriers were low perception of usefulness, no awareness of existence, lack of resources, or inadequateness in terms of outdated material or being in an immature state. We interpreted these problems as symptoms of a lack of a proper dissemination plan in the project.

A threat to validity arises from the low number of respondents to the survey and the focus in only one research project, whereby this study is limited in completeness and generalizability.

In our future work, we therefore plan to conduct a larger study in successor projects of SPES_XT to understand the real implication of these findings for the transmission of the knowledge and techniques developed into practice. In such an attempt, we will look at differences between academic and industry partners as well as a segmentation of participants by roles (e.g., managers, developers, architects, researchers).

ACKNOWLEDGMENT

We thank Michaela Tiessler for her contribution in the implementation of the survey. Also, we are thankful to all SPES_XT partners who participated to the survey.